\documentclass[12pt,final]{article}
\usepackage{graphicx}
\usepackage{epsfig}
\usepackage{dcolumn}
\usepackage{bm}

\usepackage{amssymb,amsmath}
\usepackage{gensymb}
\usepackage{multirow}
\usepackage{booktabs}
\makeatletter
\def\@biblabel#1{(#1)}
\makeatother
\usepackage{setspace}

\usepackage{booktabs,rotating}
\usepackage{fancyhdr}
\usepackage{booktabs,caption}
\usepackage[flushleft]{threeparttable}
\usepackage{enumerate,subfigure,tabularx,longtable}
\usepackage{longtable}
\usepackage[utf8x]{inputenc}
\UseRawInputEncoding
\usepackage{gensymb}
\usepackage {threeparttable} 

\newcommand{\et}{\textit{et al.}}

\newcommand{\comment}[1]{}

\usepackage{lineno}   

\def\gsim {\mbox{\hbox{ \lower-.6ex\hbox{$>$}
\kern-1.12em \lower.5ex\hbox{$\sim$}\kern+.35em}}}
\def\lsim {\mbox{\hbox{ \lower-.6ex\hbox{$<$}
\kern-1.12em \lower.5ex\hbox{$\sim$}\kern+.35em}}}
\makeatletter
\let\@fnsymbol\@arabic
\makeatother

\begin{document}


\title{\vspace{-2.0cm}
	\LARGE \textbf{Accounting for Nanofilm Contributions in Interfacial Free Energy Calculations Using Classical Density Functional Theory}}

	\author{Yafan Yang$^{\dag,\ref{fn:1}*}$, Zufeng Zuo$^{\dag}$, Xingyu Zhao$^{\ddag}$,\\ Shuyu Sun$^\S$, and Denvid Lau$^{\ddag,\ref{fn:1}*}$ \\
		\\[-15pt] 
		\small  $^{\dag}$State Key Laboratory of Intelligent Construction and Healthy Operation \\
		\small  and Maintenance of Deep Underground Engineering, \\
		\small  China University of Mining and Technology, 
		\small  Xuzhou, Jiangsu, China. \\
		\small $^\ddag$Department of Architecture and Civil Engineering, \\
		\small  City University of Hong Kong, 
		\small  Hong Kong, China. \\
		\small $^\S$School of Mathematical Sciences, Tongji University, Shanghai, China.\\
	}

\date{\today}
\maketitle

\footnotetext{\label{fn:1}$^*$ To whom correspondence should be addressed, e-mails: yafan.yang@cumt.edu.cn; denvid.lau@cityu.edu.hk.}

\newpage
\begin{abstract}
Fluid nanofilms play a fundamental role in nanoscale interfacial thermodynamics, yet their treatment in classical density functional theory (cDFT) remains incomplete. We investigate nanofilm interfacial properties using a cDFT framework built upon the perturbed-chain statistical associating fluid theory (PC-SAFT) for both fluid-fluid and fluid-solid interfacial systems. Comparison with molecular simulations shows that the long-standing numerical discrepancies regarding free-standing nanofilms persist in the PC-SAFT functional predictions, supporting the view that they stem from thermal capillary-wave fluctuations rather than deficiencies of the density functional itself, as such fluctuations are inherently neglected in the mean-field approximation adopted by standard cDFT.
Importantly, we establish a thermodynamically consistent framework for interfacial free energy (IFE) calculation, which explicitly incorporates nanofilm thermodynamic contributions and redefines the effective fluid volume by excluding the solid-phase region. The proposed method exhibits excellent consistency with another method based on the relationship between IFE and disjoining pressure for fluid systems, while both fluid-fluid and fluid-solid IFEs differ substantially from those predicted by the conventional method. 
We find that neglecting nanofilm contributions induces prominent size-dependent variations in the IFE and contact angle of hemicylindrical droplets, which is inconsistent with the extensive literature.
Different methods also lead to opposite signs of the line tension for a hemispherical argon nanodroplet on a strongly lyophobic surface. The proposed framework provides a unified molecular-level basis for understanding interfacial processes involving nanofilms, including wetting, nucleation, adsorption, and other phenomena that rely on the accurate evaluation of IFEs.
\end{abstract}

\clearpage

\newpage
\section{Introduction}

Nanofilms are ubiquitous in a wide range of natural phenomena and engineering applications, including bubble and droplet coalescence, wetting, lubrication, flotation, and colloidal stability~\cite{israelachvili2011intermolecular,saramago2010thin}. As illustrated in Fig. 1, a liquid-like nanofilm may form between approaching gas bubbles during bubble coalescence or between a gas phase and a strongly lyophilic solid during wetting, whereas a gas-like nanofilm can exist between approaching liquid droplets or between a liquid and a strongly lyophobic substrate. 
These nanofilms govern film rupture, phase coalescence, and dynamic wetting, and also play a central role in determining interfacial thermodynamic properties, including disjoining pressure, interfacial free energy (IFE), contact angle (CA), and line tension~\cite{oron1997long,nikolov2019structure}. In particular, the magnitude and even the sign of the line tension remain subjects of longstanding debate, with reported values spanning several orders of magnitude in experiments, theories, and simulations~\cite{wang2024wetting}. Accurate characterization of nanofilms is therefore essential for understanding and predicting nanoscale interfacial phenomena. Because the film thickness is comparable to the range of intermolecular interactions, molecular-level approaches are generally required to describe their thermodynamic and structural properties~\cite{henderson2005statistical,bhatt2002molecular}.

Experimental characterization of nanofilms remains difficult because their thickness is typically on the order of only a few nanometers. In addition to the limited spatial resolution of experimental techniques, free liquid nanofilms become increasingly susceptible to long-wavelength thermal fluctuations as their thickness decreases, leading to spontaneous rupture before equilibrium can be established~\cite{bhatt2002molecular,macdowell2014disjoining}. Consequently, experimental studies have mainly focused on relatively stable surfactant-stabilized aqueous films~\cite{bergeron1992equilibrium,karraker2002disjoining,stubenrauch2003disjoining} or equilibrium wetting films formed on solid substrates~\cite{sharma1993equilibrium,diakova2003thin,huerre2017laplace,zou2021disjoining}.

Molecular simulations can overcome the limitations by suppressing long-wavelength fluctuations through the finite simulation domain, thereby enabling direct investigation of metastable nanofilms, including their microscopic structure, thermodynamic properties, and stability.
Molecular simulations have been extensively employed to characterize density profiles, disjoining pressure isotherms, film stability, and the structural forces governing nanoscale interfaces~\cite{bhatt2002molecular,bhatt2004monte,peng2015methodology,yang2026resolving,jianzhou2026estimating}.


Classical density functional theory (cDFT) provides a molecular-level description of inhomogeneous fluids by directly minimizing the free-energy functional, offering a computationally efficient alternative to molecular simulations. In particular, recent developments based on the perturbed-chain statistical associating fluid theory (PC-SAFT) equation of state (EoS) have substantially improved the predictive capability and transferability of cDFT~\cite{sauer2017classical}. Extensive validations against Monte Carlo (MC) and molecular dynamics (MD) simulations have demonstrated that PC-SAFT cDFT accurately predicts density profiles, adsorption isotherms, surface tensions, and equilibrium CAs over a wide range of realistic fluid-fluid and fluid-solid systems~\cite{sauer2017classical,sauer2018prediction,camacho2019gas,sang2024investigating,thiele2026efficient}. These studies establish PC-SAFT cDFT as a computationally efficient molecular theory capable of achieving high accuracy for a broad range of nanoscale interfacial phenomena.

Nevertheless, previous applications of cDFT to nanofilms have produced inconsistent agreement with molecular simulations~\cite{bhatt2002molecular,yu2009novel}. Yu~\cite{yu2009novel} developed a weighted-density functional theory to investigate adsorption, fluid-solid interfacial tensions, and disjoining pressures of simple liquid films on planar substrates. The predicted disjoining pressures showed excellent agreement with molecular simulation results~\cite{han2008disjoining} for neon films adsorbed on graphite surfaces. In contrast, Bhatt \textit{et al.}~\cite{bhatt2002molecular} reported substantial discrepancies between perturbation-theory-based cDFT and molecular simulations for free-standing Lennard-Jones nanofilms, with predicted disjoining pressures differing by several-fold for thin films. They attributed these discrepancies to deficiencies of the perturbation-theory functional, arguing that the approximate pair correlation function employed in the density functional failed to accurately capture the fluid structure. These contrasting findings raise the question of whether the discrepancies originate from limitations of the density functional itself or from other aspects of the theoretical framework. Motivated by this unresolved issue, the present work re-examines the consistency between molecular simulations and cDFT using the more advanced PC-SAFT cDFT framework.

On the other hand, in cDFT, the IFE of planar interfaces is commonly evaluated from the excess grand potential of the inhomogeneous system according to~\cite{sauer2017classical,sauer2018prediction,camacho2019gas,zhou2013line,yatsyshin2021surface}:
\begin{equation}
	\label{eq:m3}
	\gamma=\frac{\Omega + PV}{A},
\end{equation}
where $\Omega$ is the grand potential of the inhomogeneous system, $P$ is the bulk fluid pressure, $V$ is the fluid volume, and $A$ is the interfacial area.

This expression is valid for macroscopic planar interfaces in the presence of nanofilms, where the adjoining bulk phase is at the same pressure as that of the film region. However, its applicability to nanoscale systems has received little attention. In particular, the pressure within a nanofilm generally differs from that of the adjoining bulk phase owing to the presence of disjoining pressure, suggesting that Eq.~\ref{eq:m3} may not correctly account for the free-energy contribution of the nanofilm. Despite this potential inconsistency, the same expression has been widely adopted to evaluate the IFEs of nanoscale droplets and to predict CAs and line tensions~\cite{zhou2013line,yatsyshin2021surface}.

A second issue concerns the definition of the fluid volume, $V$, in Eq.~\ref{eq:m3}. For fluid-fluid interfaces, $V$ is uniquely defined. In contrast, for fluid-solid interfaces, the presence of an external potential makes the definition of the fluid volume ambiguous because the solid phase has no sharp physical boundary. For macroscopic droplets, whether the volume occupied by the solid is included or excluded has little influence on the predicted CA because of the small volume of the solid part, and the resulting error also largely cancels when evaluating the difference between the liquid-solid and gas-solid IFEs. At the nanoscale, however, this cancellation is no longer guaranteed, and the associated error may become significant.

Overall, this work has two primary objectives. First, we systematically assess the capability of the PC-SAFT cDFT framework to quantitatively predict the structural and thermodynamic properties of free-standing nanofilms through a direct comparison with molecular simulation results. Second, and more importantly, we revisit the thermodynamic framework for calculating IFEs in cDFT, with the aim of deriving alternative formulations that explicitly account for nanofilm contributions and quantify the impact of different fluid-volume definitions in the presence of an external potential. The proposed formulations are applied to both fluid-fluid and fluid-solid interfaces. 

By executing these objectives, we clarify the intrinsic role of thermal capillary-wave fluctuations in reconciling the long-standing quantitative discrepancies between cDFT and MD profiles for free-standing films. Moreover, by addressing critical deficiencies in traditional cDFT formulations, specifically where the neglect of nanoscopic fluid film contributions and solid-volume ambiguities occurs, the refined thermodynamic framework successfully eliminates unphysical artifacts in contact angle predictions for nanoscale droplets.
Consequently, this framework provides a robust molecular-level computational methodology that can help address the long-standing debate over the magnitude and sign of line tension.
Ultimately, this work provides new molecular-level insights into nanofilm thermodynamics and establishes a rigorous theoretical framework for accurately evaluating IFEs, thereby improving the predictive capability of molecular theories for nanoscale wetting and related interfacial phenomena.

\section{Method}

\subsection{cDFT}
cDFT was employed to investigate inhomogeneous fluids. Two types of systems were considered in the present study using one-dimensional cDFT: fluid systems and fluid-solid systems. Calculations for the former were carried out with a modified version of SurfPack from the ThermoTools project~\cite{SurfPack2024}, while those for the latter were performed using the FeO\textsubscript{s} framework~\cite{rehner2023feos}.

At prescribed temperature $T$, volume $V$, and chemical potential $\mu$, the equilibrium density profile $\rho(\mathbf{r})$ with $\mathbf{r}$ as a spatial vector is obtained by minimizing the grand potential functional $\Omega_V$~\cite{roth2002fundamental}:
\begin{equation}
	\Omega_V[\rho(\mathbf{r})] = F[\rho(\mathbf{r})] + \int \rho(\mathbf{r}) \{V^{\mathrm{ext}}(\mathbf{r})-\mu\} d\mathbf{r}.
\end{equation}

We adopt the Helmholtz free energy functional $F$ model from the PC-SAFT equation of state~\cite{sauer2017classical}:
\begin{equation}
	F=F^{id}+F^{res} =F^{id}+F^{hs}[\rho(\mathbf{r})]
	+F^{hc}[\rho(\mathbf{r})]
	+F^{disp}[\rho(\mathbf{r})]
	+F^{assoc}[\rho(\mathbf{r})],
\end{equation}
where $F^{\mathrm{id}}$ is the ideal gas contribution and $F^{\mathrm{res}}$ is the residual Helmholtz free energy, which comprises the hard-sphere $F^{\mathrm{hs}}$~\cite{roth2002fundamental,yu2002structures}, chain $F^{\mathrm{hc}}$~\cite{tripathi2005microstructure,tripathi2005microstructure2}, dispersion $F^{\mathrm{disp}}$~\cite{sauer2017classical}, and association $F^{\mathrm{assoc}}$~\cite{yu2002fundamental} contributions.

The influence of the solid surface on the fluid is incorporated through the external potential, $V^{\mathrm{ext}}(\mathbf{r})$, which describes the interaction acting on a fluid interaction site at position $\mathbf{r}$. Within the grand canonical framework, the equilibrium state corresponds to the minimum of the grand potential functional, $\Omega_V[\rho(\mathbf{r})]$. Consequently, the equilibrium grand potential is given by $\Omega=\min\Omega_V$, and the associated density profile satisfies the Euler--Lagrange equation obtained from the stationary condition~\cite{roth2002fundamental}:
\begin{equation}
	\frac{\delta\Omega_V}{\delta\rho(\mathbf{r})}=0.
\end{equation}

The equilibrium density of the grand canonical follows~\cite{roth2002fundamental}:
\begin{equation}
	\rho(\mathbf{r})=
	\frac{1}{\Lambda^{3}}
	\exp\left(
	\beta\mu
	-\frac{\delta \beta F^{\mathrm{res}}}{\delta \rho(\mathbf{r})}
	-\beta V^{\mathrm{ext}}(\mathbf{r})
	\right),
	\label{eq:rho}
\end{equation}
where $\beta=1/(kT)$, with $k$ denoting the Boltzmann constant, and $\Lambda$ representing the thermal de Broglie wavelength.

For calculations of nanofilm properties in the fluid systems, the grand canonical ensemble DFT described above converges to different equilibrium solutions depending on the initial density profile~\cite{bhatt2002molecular}. Bhatt \textit{et al.}~\cite{bhatt2002molecular} addressed this issue by testing multiple initial density profiles and identifying the solution that conserves the total number of particles. This procedure can be regarded as an alternative way of minimizing the Helmholtz free energy, $F$, under the constraint of a fixed particle number, $N$~\cite{bhatt2002molecular}.

In the present study, we instead employ a canonical ensemble DFT approach, following the formulation of Sauer \textit{et al.}~\cite{sauer2018prediction}. In this framework, the total number of molecules is explicitly constrained by introducing a Lagrange multiplier, thereby ensuring particle-number conservation throughout the minimization process. The total number of molecules is given by：
\begin{equation}
	N=\int \rho(\mathbf{r})\,\mathrm{d}\mathbf{r}.
\end{equation}

Substituting Eq.~(\ref{eq:rho}) into the normalization condition yields~\cite{sauer2018prediction}:
\begin{equation}
	N=
	\frac{1}{\Lambda^{3}}
	\exp(\beta\mu)
	\int
	\exp\left(
	-\frac{\delta\beta F^{\mathrm{res}}}
	{\delta\rho(\mathbf{r})}
	-\beta V^{\mathrm{ext}}(\mathbf{r})
	\right)
	\mathrm{d}\mathbf{r}.
\end{equation}

Eliminating the chemical potential from the above expression leads to the Euler--Lagrange equation for the particle-number-constrained formulation~\cite{sauer2018prediction}:
\begin{equation}
	\rho(\mathbf{r})
	=
	\frac{
		N
		\exp\left(
		-\dfrac{\delta\beta F^{\mathrm{res}}}
		{\delta\rho(\mathbf{r})}
		-\beta V^{\mathrm{ext}}(\mathbf{r})
		\right)
	}{
		\displaystyle
		\int
		\exp\left(
		-\dfrac{\delta\beta F^{\mathrm{res}}}
		{\delta\rho(\mathbf{r})}
		-\beta V^{\mathrm{ext}}(\mathbf{r})
		\right)
		\mathrm{d}\mathbf{r}
	}.
\end{equation}

The resulting equation is solved iteratively using a damped Picard scheme in SurfPack.

\subsection{Interfacial properties in fluid systems}

Pure argon and water systems were investigated in this work. Argon was modeled using the Lennard-Jones PC-SAFT parameters reported in Ref. \cite{sauer2018prediction}, while water was described by the PC-SAFT model with the 4C association scheme proposed in Ref. \cite{liang2014modeling}.
The fluid systems considered here involve two types of nanofilms: liquid-like films and gas-like films. A liquid-like film is bounded by two bulk vapor phases, whereas a gas-like film is bounded by two bulk liquid phases (see Figs. \ref{fig:z1}a and b). 

The cDFT calculation of the nanofilm system is analogous to that used in molecular simulations~\cite{bhatt2002molecular,yang2026resolving}. Only the nanofilm and the adjacent bulk phase are explicitly included in the calculation, while the other bulk phase, which is connected to the nanofilm through the meniscus in the actual system, is not modeled explicitly. Instead, its thermodynamic properties are estimated using the PC-SAFT EoS. 
The nanofilm is assumed to have a planar interface; therefore, the one-dimensional cDFT calculations are performed in Cartesian coordinates.
The film properties were studied as a function of film thickness $h$. In the limit of infinite film thickness, the density in the central region approaches the corresponding bulk phase density, thereby recovering the bulk thermodynamic properties. 

Three methods are compared for calculating the IFEs of the two interfaces bounding the nanofilm. Note that the IFE is equivalent to the surface tension for the planar fluid interfaces considered in this study.

Method I is based on Gibbs' concept of the dividing surface~\cite{gibbs2013collected}. Following a similar derivation of the IFE for droplet systems~\cite{rehner2018surface}, we construct a hypothetical reference system that shares the same $T$, $V$, and $\mu$ as the original fluid film system. This reference system comprises homogeneous liquid ($L$) and gas ($G$) phases separated by two infinitely thin interfaces ($\sigma$). Accordingly, any extensive thermodynamic quantity $X$ can be written as the sum of the contributions from these three regions,
\begin{equation}
	\label{XXX}
	X = X_{L} + X_{G} + X_{\sigma}.
\end{equation}

The interfacial region is defined to have zero volume, i.e.,
\begin{equation}
	V_{\sigma}=0.
\end{equation}

Accordingly, the total number of molecules can be written as
\begin{equation}
	N = N_{L} + N_{G} + N_{\sigma}
	= \rho_{L}V_{L} + \rho_{G}V_{G} + N_{\sigma},
\end{equation}
where $\rho_{L}$ and $\rho_{G}$ denote the bulk liquid and vapor densities at the specified $T$ and $\mu$, respectively. The equimolar dividing surface is then defined as the dividing surface for which the interfacial excess number of molecules vanishes, $i.e.$, $N_{\sigma}=0$. 
For planar nanofilms, the film thickness, $h$, is defined as the distance between the two equimolar dividing surfaces~\cite{ivanov1975thermodynamics2,henderson2005statistical,yang2026resolving}. 
Therefore, the thickness of a liquid-like film is
\begin{equation}
	h=\frac{N/A-\rho_{G}L}{\rho_{L}-\rho_{G}},
\end{equation}
whereas that of a gas-like film is
\begin{equation}
	h=\frac{N/A-\rho_{L}L}{\rho_{G}-\rho_{L}},
\end{equation}
where $L$ is the domain length normal to the interface.

Eq. \ref{XXX} can be applied to the grand potential of the system:
\begin{equation}
	\Omega = \Omega_L + \Omega_G + \Omega_{\sigma} = -P_LV_L-P_GV_G+\Omega_{\sigma}.
\end{equation}

The IFE for one surface of the film is:
\begin{equation}
	\label{eq:m1}
	\gamma = \cfrac{\Omega_{\sigma}}{2A} = \cfrac{\Omega+ P_LV_L + P_GV_G}{2A}.
\end{equation}

For an interface separating two bulk phases, the surface of tension used to define $V_L$ and $V_G$ in Eq.~\ref{eq:m1} does not generally coincide with the Gibbs dividing surface~\cite{lovett1997molecular,rehner2018surface}. In contrast, nanofilms studied here lack a bulk-like interior, and the Gibbs dividing surface preserves the thermodynamic relation between the disjoining pressure and the surface tension, as discussed below. Accordingly, the Gibbs dividing surface is adopted to define $V_L$ and $V_G$ in Eq.~\ref{eq:m1}.

Method II is based on the relation between the disjoining pressure of the film and the IFE. Applying Eq. \ref{XXX} to the Helmholtz free energy differential of a pure system composing two bulk phases and a film gives~\cite{toshev1975thermodynamics}:
\begin{equation}
	2d\gamma = -\tilde{s}_{\sigma} dT-\Pi dh - \tilde{n}_{\sigma} d\mu.
\end{equation}
where $\tilde{s}_{\sigma} = S_{\sigma}/A$ is the specific excess entropy (defined as the excess entropy $S_{\sigma}$ per unit area), $\Pi$ is the disjoining pressure of the film, and $\tilde{n}_{\sigma} = N_{\sigma}/A$ is the specific excess number of molecules. By adopting Gibbs dividing surface, for which $\tilde{n}_{\sigma} = 0$, and considering isothermal equilibrium ($dT = 0$), the differential relation reduces to~\cite{ivanov1975thermodynamics2}:
\begin{equation}
	2d\gamma = -\Pi dh.
\end{equation}

Integrating the above equation from $h$ to $\infty$ yields the following expression for IFE:
\begin{equation}
	\label{eq:m2}
	\gamma = \gamma_{\infty} + \frac{1}{2}\int_{h}^{\infty}\Pi\,dh,
\end{equation}
where $\gamma_{\infty}$ is the surface tension of two bulk phases, corresponding to the limit where the film is sufficiently thick that its central region becomes bulk-like.

Calculating the IFE using Eq.~\ref{eq:m2} requires $\Pi$. From canonical DFT calculations, the pressure of one bulk phase is readily available, while the pressure of the other bulk phase is determined from the EoS at the same $T$ and $\mu$. $\Pi$ is then evaluated from its mechanical definition as $\Pi = P_G - P_L$ for liquid-like films and $\Pi = P_L - P_G$ for gas-like films.

Method III uses Eq. \ref{eq:m3} without considering the film contribution for comparison.

\subsection{Interfacial properties in fluid-solid system}

The interfacial properties of argon droplets on a solid substrate were investigated using a Lennard-Jones 9-3 external potential~\cite{sauer2018prediction}:
\begin{equation}
	V^{\mathrm{ext}}(z)=
	\frac{\epsilon_{sf}\sigma_{sf}^{3}\rho_s}{45}
	\left[
	2\left(\frac{\sigma_{sf}}{z}\right)^9
	-15\left(\frac{\sigma_{sf}}{z}\right)^3
	\right],
\end{equation}
where $\epsilon_{sf}$ and $\sigma_{sf}$ are the solid-fluid  energy and size parameters, respectively, $\rho_s = 0.1137\ \mathrm{\AA}^{-3}$ is the number density of the solid, and $z$ is the distance from the substrate. The solid-fluid interaction energy is given by $\epsilon_{sf} = \xi \epsilon_{ff}$, where $\epsilon_{ff}/k = 119.86$ K is the fluid-fluid interaction energy parameter of argon~\cite{sauer2017classical}. The interaction parameter was fixed at $\xi = 0.25$, corresponding to a strongly lyophobic surface~\cite{sauer2018prediction}. The size parameter was taken as $\sigma_{sf} = \sigma_{ff} = 3.405\ \mathrm{\AA}$~\cite{sauer2017classical}.

1D cDFT is usually used to estimate the IFE of each interface to calculate the CA of the macroscopic droplet $\theta_{\mathrm{Y}}$ on a solid surface using Young's equation:
\begin{equation}
	\label{eq:Young}
	\gamma_{SG}-\gamma_{SL}-\gamma_{GL} \cdot cos \theta_{\mathrm{Y}}=0,
\end{equation}
where $\gamma_{SG}$, $\gamma_{SL}$, and $\gamma_{GL}$ are the solid-gas, solid-liquid, and gas-liquid IFEs, respectively. It should be noted that the IFE of a solid-fluid interface should include a contribution from the solid phase. In the present work, however, the solid is represented by an external potential, and this contribution is therefore omitted. Consequently, the calculated solid-fluid IFEs are relative rather than absolute values and may result in negative values~\cite{yang2025estimating}.

As the droplet size decreases, the pressure difference between the droplet and bulk gas arises from the combined effects of the capillary pressure associated with the curved gas-liquid interface, the disjoining pressure originating from the thin film near the substrate, and the line tension acting at the three-phase contact line. These effects are incorporated in the augmented Young's equation~\cite{drelich1996significance}:
\begin{equation}
	\label{eq:augYoung}
	\gamma_{SG}-\gamma_{SL}-\gamma_{GL}\cos\theta-\frac{\tau}{R_e\sin\theta}=0,
\end{equation}
where $\theta$ is the microscopic CA determined from the density profile of the hemispherical droplet, $R_e$ is the radius of the droplet defined by the Gibbs dividing surface, and $\tau$ is the line tension.

Substituting Eq. \ref{eq:Young} into Eq. \ref{eq:augYoung} yields~\cite{drelich1996significance}:
\begin{equation}
	\label{eq:augYoung2}
	\cos \theta_{\mathrm{Y}}-\cos\theta-\frac{\tau}{\gamma_{GL}R_e\sin\theta}=0.
\end{equation}

In this work, 1D cDFT is employed to calculate the $\gamma_{SG}$, $\gamma_{SL}$, and $\gamma_{GL}$ for a hemispherical argon droplet interacting with the external potential, $V^{\mathrm{ext}}(z)$. The same system was previously studied by Sauer \et~\cite{sauer2018prediction}, who determined the $\theta$ from droplet density profiles predicted by three-dimensional molecular MC simulations and two-dimensional cDFT calculations in cylindrical coordinates. This enables the evaluation of the line tension, $\tau$.

According to the chemical equilibrium condition, the thermodynamic state of the three-phase argon system in the presence of the external potential is specified by a single chemical potential. Unlike the macroscopic limit, the equilibrium chemical potential of a nanoscale droplet depends on its size, giving rise to the pressure difference between the liquid droplet and the surrounding gas. To enable comparison with the reported CA of hemispherical droplets, the solid-fluid IFEs were first calculated over a range of chemical potentials, and the resulting relationship between the pressure difference and the IFEs was established. Although the chemical potential of the droplet was not reported in Ref.~\cite{sauer2018prediction}, the Gibbs dividing surface radius, $R_e$, can be estimated from the reported simulation details. The corresponding capillary pressure is then determined from the relationship between $R_e$ and the pressure difference obtained from calculations of spherical argon droplets. Finally, this pressure difference is used together with the established pressure difference-IFE relationship to determine the corresponding solid-fluid IFEs.

The gas-liquid IFE, $\gamma_{GL}$, of spherical argon droplets was determined using one-dimensional cDFT in spherical coordinates following the method described in Ref.~\cite{rehner2018surface}.

The calculation of the solid-fluid IFEs is more complicated due to the presence of the solid substrate. In a pure fluid system, the fluid volume is well defined. However, when an external potential is used to represent the solid, there is no sharp boundary between the fluid and the substrate because the fluid density decays continuously into the repulsive region of the external potential. Consequently, the fluid volume appearing in Eq.~\ref{eq:m3} is not uniquely defined. The total system volume has been widely adopted in the literature for IFE calculation in the presence of an external potential.

For macroscopic droplets, this approximation introduces negligible error because the pressure difference between the liquid and gas phases vanishes and the solid volume is nearly identical for the solid-liquid and solid-vapor interfaces. Consequently, the contribution of the solid volume effectively cancels out in the evaluation of macroscopic CAs. For nanodroplets, however, the finite capillary pressure makes the volume definition much more important, and including the solid volume may introduce significant errors in the calculated IFEs. To investigate this effect, three different methods were considered.

Method A extends Method I of the fluid systems by accounting for the presence of the nanofilm while excluding the solid volume. 
Note that, unlike nanofilms in fluid systems, nanofilms on solid substrates are inherently asymmetric. Consequently, the total IFE is calculated without the factor of 2 in Eq.~\ref{eq:m2}. For example, the IFE of a solid-film-liquid interface considers contributions from the liquid-film and the film-solid interactions.
The fluid volume is evaluated directly from the density profile and is defined as the region where the fluid density exceeds a small threshold ($\rho > 10^{-20}\ \mathrm{\AA}^{-3}$). This criterion slightly overestimates the fluid volume because the fluid density extends slightly into the repulsive region of the external potential. However, the resulting error from this simple approximation is expected to be negligible since the substrate is highly lyophobic and fluid penetration is minimal.

Method B is based on Method III of the fluid systems. The solid volume is excluded using the same volume definition as in Method A, but no correction is made for the nanofilm.

Method C directly applies Method III of the fluid systems without accounting for either the nanofilm or the solid volume. Consequently, the total system volume is used in the IFE calculation.

\section{Results and Discussion}

\subsection{Fluid systems}

Fig. \ref{fig:z2} compares the IFEs calculated using different methods for liquid- and gas-like films in pure argon and water systems. The calculated bulk surface tension values are in good agreement with the experimental data~\cite{linstrom2001nist}.
In general, the IFE increases with film thickness and approaches the bulk surface tension in the limit of large film thickness (large $h$).
An exception is observed for the water gas film, where the IFE exhibits an oscillatory dependence on film thickness.
As $h$ decreases, the IFE initially decreases slightly, followed by a moderate increase and then a sharp decline.
Similar oscillatory behavior has previously been reported in theoretical studies of liquid films on solid substrates~\cite{macdowell2014disjoining,macdowell2013capillary} and experimental measurements of surfactant-stabilized free-standing water films~\cite{bergeron1992equilibrium}.
The agreement between our predictions and these previous observations suggests that such oscillatory behavior is a more general characteristic of nanofilms.
It originates from the structural ordering of molecules within the film: as the film becomes sufficiently thin, the diffuse density profiles associated with the two fluid interfaces increasingly overlap, resulting in alternating molecular packing configurations. Consequently, the excess free energy oscillates with film thickness, giving rise to the oscillatory variation of the IFE.

Importantly, excellent agreement is observed between Methods I and II, with a maximum deviation of only 0.2\%, confirming the thermodynamic consistency of the two approaches. In contrast, although Method III correctly recovers the bulk surface tension in the limit of large film thicknesses, it predicts a premature reduction in the IFE as the film becomes thinner, with significant deviations emerging at much larger film thicknesses than those obtained using Methods I and II. These results highlight the crucial role of explicitly accounting for nanofilm thermodynamics in the accurate evaluation of IFE in thin-film systems.

The corresponding disjoining pressure is shown in Fig. \ref{fig:z3}.
In general, the disjoining pressure increases with decreasing film thickness.
An exception is again observed for the water-gas film, where $\Pi$ exhibits an oscillatory dependence on film thickness, mirroring the corresponding oscillatory behavior of the IFE.
The MD results for the liquid films are also included for comparison~\cite{jianzhou2026estimating}, whereas no corresponding MD data are available for the gas films.
For clarity, only one set of MD results is presented, as the different MD methods yield nearly identical predictions~\cite{bhatt2002molecular,peng2015methodology,yang2026resolving,jianzhou2026estimating}.

The comparison between cDFT and MD indicates that the advanced PC-SAFT density functional does not resolve the discrepancy previously reported by Bhatt \textit{et al.}~\cite{bhatt2002molecular}. Specifically, the MD results exhibit a decrease in the IFE at considerably larger film thicknesses as the film becomes thinner, whereas the cDFT predictions remain close to the bulk value until much smaller film thicknesses. This finding suggests that the discrepancy is unlikely to originate solely from deficiencies of the perturbation-theory functional employed by Bhatt \textit{et al.}~\cite{bhatt2002molecular}, since a substantially more sophisticated PC-SAFT density functional leads to the same qualitative behavior.

Instead, the discrepancy appears to be associated with the fundamental differences between free-standing nanofilms and supported films, as well as with the intrinsic limitations of the mean-field description adopted in classical DFT. Previous studies of adsorbed liquid films on solid substrates have demonstrated excellent agreement between cDFT and molecular simulations for density profiles and disjoining pressures~\cite{yu2009novel,han2008disjoining}. In such systems, the fluid structure is primarily governed by the external solid potential, which strongly constrains the interface and suppresses long-wavelength interfacial fluctuations. Consequently, the intrinsic interfacial structure predicted by mean-field cDFT closely resembles the thermally averaged structure sampled in molecular simulations~\cite{macdowell2014disjoining}. Another factor contributing to this difference is the low temperature adopted in the fluid-solid system~\cite{yu2009novel,han2008disjoining}, which further suppresses thermal capillary-wave fluctuations.

In contrast, a free-standing nanofilm consists of two diffuse liquid-gas interfaces interacting in the absence of any external confining potential. The film free energy is therefore determined by the overlap of the two intrinsic interfaces, making the IFE and disjoining pressure substantially more sensitive to subtle variations in the interfacial structure than in fluid-solid systems. Moreover, classical DFT determines the equilibrium density profile by minimizing the free-energy functional and therefore predicts the intrinsic interfacial free energy, whereas molecular simulations naturally incorporate thermal capillary-wave fluctuations that contribute an additional entropic component to the free energy~\cite{macdowell2014disjoining,evans1979nature}. Because both interfaces are free to fluctuate, these fluctuation effects are expected to become increasingly important as the two interfaces begin to overlap, causing the effective interaction between the interfaces to emerge at larger film thicknesses than predicted by the intrinsic mean-field description. Consequently, the reduction of the IFE is observed at considerably larger film thicknesses in MD than in cDFT.

These observations imply that improving the intrinsic free-energy functional alone may not be sufficient to achieve quantitative agreement between cDFT and molecular simulations for free-standing nanofilms. Instead, the discrepancy may reflect a more fundamental limitation of the mean-field framework, which neglects thermal interfacial fluctuations that become increasingly relevant when the interaction between two free interfaces governs the thermodynamics of the system.

Fig. \ref{fig:z4} presents the density profiles extending from the center of the film ($z=0$) to the bulk region across the interface. The density profiles closely follow a hyperbolic tangent (\textit{tanh}) function, in agreement with previous simulation and theoretical studies~\cite{camacho2019gas,stephan2018vapor,yang2024interfacial}. As the film becomes thinner, the density at the film center gradually deviates from its bulk value, decreasing for liquid-like films (Figs. \ref{fig:z4}a, b, d, and e) and increasing for gas-like films (Figs. \ref{fig:z4}c and f).

A pronounced difference is observed between the density profiles predicted by cDFT and those obtained from MD~\cite{yang2026resolving} for liquid-like films. Although the film thicknesses are not identical, comparison over similar thickness ranges clearly shows that the cDFT profiles (Figs. \ref{fig:z4}a and d) exhibit substantially sharper liquid-gas interfaces than the corresponding MD profiles (Figs. \ref{fig:z4}b and e). In addition, the MD density profiles evolve much more rapidly as the film becomes thinner. The deviation of the central density from its bulk value appears at considerably larger film thicknesses than predicted by cDFT and becomes increasingly pronounced as the film approaches rupture. By contrast, the cDFT density profiles remain nearly unchanged over a broad range of film thicknesses, with noticeable density depletion occurring only at much smaller film thicknesses. It is also noteworthy that the interfacial width in the MD profiles increases progressively with decreasing film thickness, whereas the cDFT profiles retain nearly the same interfacial sharpness throughout most of the thickness range.

These observations are fully consistent with the IFE/$\Pi$ behavior discussed above. The broader interfaces and earlier density depletion predicted by MD indicate that the overlap between the two diffuse liquid-gas interfaces becomes significant at considerably larger film thicknesses than predicted by cDFT, leading to an earlier reduction of the IFE/$\Pi$. A plausible explanation is that classical DFT, as a mean-field theory, predicts the intrinsic density profile by minimizing the free-energy functional, whereas MD naturally incorporates thermal capillary-wave fluctuations~\cite{macdowell2014disjoining,evans1979nature,stephan2018vapor}. Such fluctuations broaden the apparent liquid-gas interface and enhance the effective overlap between the two free interfaces, effects that are expected to become increasingly important as the film approaches rupture. Consequently, the density profile analysis provides further evidence that the systematic discrepancy between cDFT and MD for free-standing nanofilms is unlikely to arise solely from the choice of density functional, but may instead reflect a more fundamental limitation of the intrinsic mean-field description when thermal interfacial fluctuations become significant.

\subsection{Fluid-solid system}
Fig. \ref{fig:z5} presents the density profiles of argon in the liquid-film-solid interfacial region at different disjoining pressures.
The solid volume is marked as the gray-shaded region. 
The fluid density distributions exhibit a clear spatial organization adjacent to the solid substrate, with pronounced layering induced by fluid-solid interactions. 
The Gibbs dividing surfaces between the gas-like film and the bulk liquid are indicated by the dotted vertical lines.
As the disjoining pressure varies, the thickness of the liquid-like film adjusts accordingly, leading to systematic changes in the density profiles. With increasing disjoining pressure, the film becomes thinner, and the two interfaces approach each other, whereas at $\Pi$ = 0 MPa, a clearer separation between bulk liquid and bulk gas regions is recovered, consistent with the classical definition of bulk phases based on the Gibbs construction.

We computed the IFE of a spherical argon droplet $\gamma_{LG}$ following the method described in Ref. \cite{rehner2018surface}. Figs. \ref{fig:z6}a and b present the resulting IFE and capillary pressure ($\Delta P$) as functions of $1/R_e$.
We obtain a similar trend of $\gamma_{LG}$ as reported in Ref. \cite{rehner2018surface}. With decreasing droplet radius, the interfacial tension first increases to a maximum and then decreases, which can be physically attributed to curvature-modified interfacial structure at intermediate radii and diffuse-interface effects at small radii.
We note that these quantities are conventionally expressed as functions of the surface of tension radius ($R_s$). However, to facilitate direct comparison with the radius $R_e$ obtained from the 2D cDFT calculations, we adopt $R_e$ as the independent variable.

In Fig. \ref{fig:z6}c, the solid-film-liquid IFE $\gamma_{SFL}$ remains positive and increases monotonically with rising disjoining pressure, indicating enhanced contractive forces under nanoconfinement.
Different methods produce markedly divergent results. 
Method A yields consistently lower values than Methods B and C, while Method C gives the highest values over the entire range. Method B remains intermediate. The three methods converge only at small disjoining pressures. The observed discrepancies indicate that method-dependent treatments of both the gas-like film IFE and the solid-volume contribution must be carefully accounted for to ensure quantitative accuracy.

The dependence of film thickness on $\Pi$ is shown in Fig. \ref{fig:z6}d. A monotonically decreasing trend is observed, indicating that increasing $\Pi$ progressively compresses the gas-like film, as evidenced by the enhanced structural ordering shown in Fig. \ref{fig:z5}.

Fig. \ref{fig:z6}e shows the solid-gas IFE $\gamma_{SG}$. In the absence of a liquid-like nanofilm at this strongly lyophobic solid-gas interface, Methods A and B yield identical results. In contrast, including the solid-volume contribution shifts $\gamma_{SG}$ from negative to positive values. Although some method-dependent differences are observed, all values are negligible compared with $\gamma_{SFL}$.

The contact angle $\theta_{\mathrm{Y}}$ estimated from Eq. \ref{eq:Young} is shown in Fig. \ref{fig:z6}f. A clear method-dependent trend is observed. Method A predicts $\theta_{\mathrm{Y}}$ to be nearly independent of $\Pi$, and thus insensitive to droplet size, whereas Methods B and C exhibit significant variations.

We argue that the CAs predicted by Method A are more consistent with previous cDFT~\cite{weijs2011origin} and MD simulation studies~\cite{weijs2011origin,silvestri2019wetting}. These studies have shown that cylindrical droplets exhibit only a weak radius dependence, whereas hemispherical droplets display a much stronger size effect. Since Young’s equation does not include line tension, the $\theta_{\mathrm{Y}}$ obtained here can be interpreted as corresponding to a cylindrical droplet with half the radius of the spherical droplet and is therefore expected to remain close to its macroscopic value until an extremely small radius. Consistently, Method A, which accounts for the gas-like nanofilm in the droplet–solid IFE, predicts CAs close to the macroscopic limit at $\Pi = 0\ \mathrm{MPa}$, while Methods B and C yield significantly larger values, in contradiction with previous observations.

Sauer \textit{et al.}~\cite{sauer2018prediction} performed combined MC simulations and cDFT calculations for the same argon-solid system. Their predicted $\theta_{\mathrm{Y}}$ at $\Pi = 0$ is 153.98$^\circ$, in excellent agreement with our value of 153.31$^\circ$. For a hemispherical droplet with the same number of particles, they reported CAs of 160.60$^\circ$ (MC) and 161.05$^\circ$ (2D cDFT). These data enable an estimation of the line tension based on the 2D cDFT results. Specifically, the droplet radius in the 2D cDFT calculation is estimated to be 36.64~\AA{} from the modeling details provided in Ref.~\cite{sauer2018prediction}. Using this radius and the corresponding disjoining pressure, we extrapolate the IFEs for each interface from Fig.~\ref{fig:z6} (see green vertical lines), and the corresponding CAs obtained from different methods are shown in Fig.~\ref{fig:z7}a.
As expected, Method A predicts a $\theta_{\mathrm{Y}}$ =154.26$^\circ$ very close to its macroscopic value, while Methods B ($\theta_{\mathrm{Y}}$ =168.88$^\circ$) and C ($\theta_{\mathrm{Y}}$ =180.00$^\circ$) predicts a much higher $\theta_{\mathrm{Y}}$ values.

The line tension is then calculated using Eq.~\ref{eq:augYoung2}, with the resulting values also included in the figure. Method A predicts a positive line tension $\tau$ = 0.77 pN, whereas Methods B ($\tau$ = -0.61 pN) and C ($\tau$ = -2.14 pN) yield negative values.
From a simple analysis of Eq.~\ref{eq:augYoung2}, the sign of the line tension is directly related to the slope of $\theta$ as a function of $1/R_e$. The data from Sauer \textit{et al.}~\cite{sauer2018prediction} indicate a positive line tension, as evidenced by the increase in CA with decreasing droplet radius (see the red and blue horizontal lines in Fig.~\ref{fig:z7}a). This observation is consistent with the prediction of Method A and supports the proposed approach for evaluating the droplet-solid IFE.

It should be emphasized that the positive line tension obtained here is not a universal characteristic of argon droplets on solid substrates. Previous studies have reported both positive~\cite{polovinkin2024contact} and negative~\cite{weijs2011origin} line tensions, depending on the interfacial interactions and substrate wettability. The key implication of the present work is therefore not the sign itself, but the importance of employing a thermodynamically consistent framework for evaluating the droplet-solid IFE, which provides a more reliable basis for determining the line tension, as illustrated in Fig.~\ref{fig:z7}b.

\section{Conclusion}
In this work, we investigated the interfacial properties of nanofilms using PC-SAFT cDFT and revisited the thermodynamic framework for calculating IFEs in nanoscale systems. 
The capability of PC-SAFT cDFT for predicting interfacial properties of free-standing nanofilms was assessed through direct comparison with molecular simulation results. Although the calculated bulk surface tensions agree well with experimental data and the proposed IFE formulations based on the grand potential and disjoining pressure exhibit excellent mutual consistency, the notable discrepancy between cDFT and MD for free-standing nanofilms remains. The advanced density functional reproduces the same qualitative deviations previously reported using perturbation-theory-based cDFT~\cite{bhatt2002molecular}, indicating that the disagreement is unlikely to originate from deficiencies of the density functional itself. Instead, the comparison of disjoining pressures and density profiles suggests that the discrepancy mainly arises from thermal capillary-wave fluctuations incorporated in molecular simulations but absent from the intrinsic mean-field description of cDFT. These observations clarify the origin of the long-standing inconsistency between cDFT and MD for free-standing nanofilms and indicate that further improvements require incorporating interfacial fluctuation effects rather than merely refining the intrinsic free-energy functional.

A thermodynamically consistent framework was developed for calculating IFEs by explicitly accounting for nanofilm contributions and resolving the ambiguity associated with the fluid-volume definition in the presence of external potentials. For fluid-fluid systems, the proposed formulations based on Gibbs dividing surfaces and the disjoining-pressure relation produce nearly identical results, confirming their thermodynamic consistency. For fluid-solid systems, neglecting nanofilm contributions or including the solid volume in the fluid volume introduces significant errors in the calculated IFEs, leading to unrealistic contact-angle predictions and even opposite signs of the predicted line tension. By contrast, the proposed formulation yields contact angles consistent with previous MC simulations and 2D cDFT calculations~\cite{sauer2018prediction} and predicts a physically reasonable positive line tension for the argon-lyophobic system.

Overall, this work establishes a thermodynamically consistent framework for evaluating the interfacial free energies of nanoscale fluid-fluid and fluid-solid systems within cDFT. By explicitly accounting for nanofilm contributions and resolving the ambiguity associated with fluid-volume definitions, the proposed methodology enables more reliable predictions of interfacial free energies, contact angles, and line tensions. More broadly, this work demonstrates that nanofilm thermodynamics should be explicitly incorporated into cDFT descriptions of nanoscale interfaces. The resulting framework provides a robust molecular-level foundation for future studies of wetting, nucleation, adsorption, confined fluids, and other interfacial phenomena where accurate evaluation of interfacial free energies is essential.


\bigskip
\bigskip
{\bf{ACKNOWLEDGMENTS\\[1ex]}}
The research is supported by the Fundamental Research Funds for the Central Universities (2025QN1175).

\bibliography{IFE_cDFT}

\clearpage

\begin{figure}[tb]
	\centering
	\includegraphics[width=0.8\textwidth]{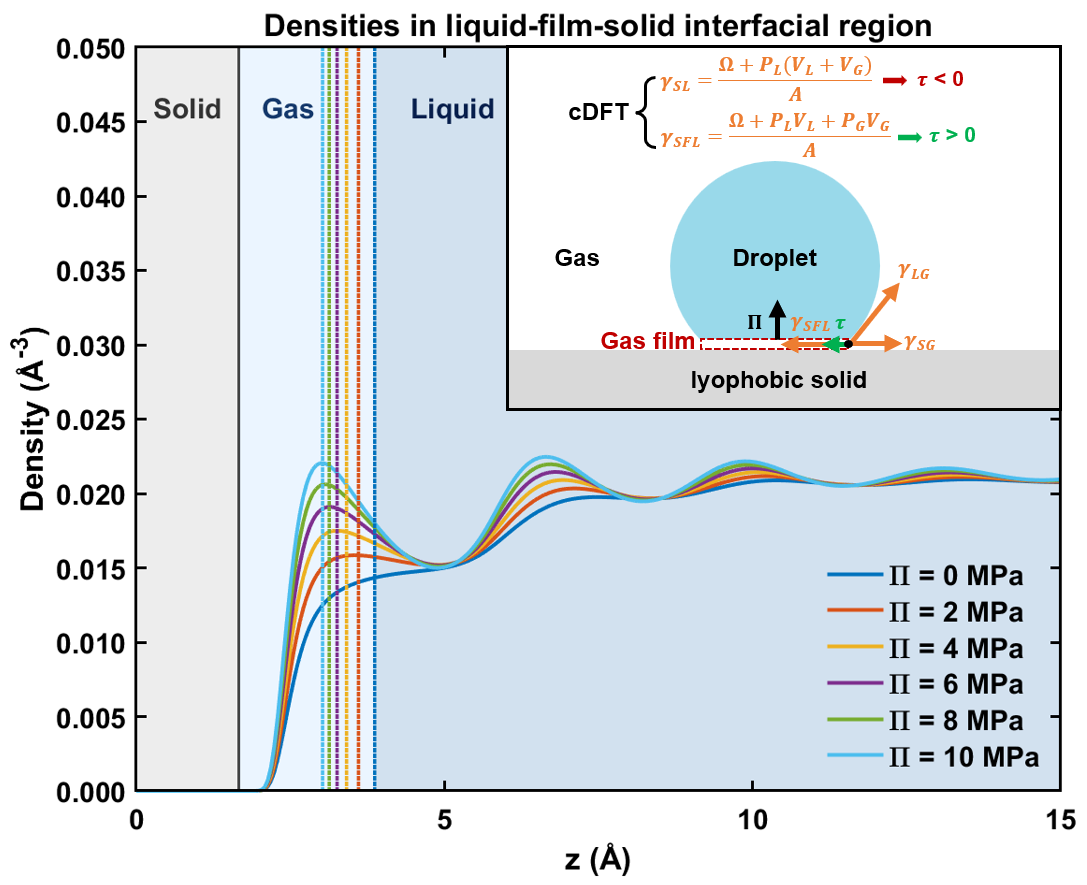}
	
	\vspace{0.5cm}
	
	{\large\bfseries Graphical Abstract\par}
	
	\vspace{1.5ex}
	
	\noindent\parbox{\textwidth}{%
		\setlength{\parindent}{2em}%
		\indent Density profiles of argon in the liquid-film-solid interfacial region at different disjoining pressures. The dotted vertical lines denote the Gibbs dividing surface for each density profile. The gray shaded region represents the solid phase. For the profile at disjoining pressure $\Pi=0$ MPa, the light blue and dark blue shaded regions indicate the bulk gas and bulk liquid phases, respectively, as defined by the Gibbs dividing surface. Inset: Schematic illustration of a gas-like nanofilm separating a hemispherical droplet from a lyophobic substrate. Different cDFT definitions of the droplet-solid interfacial free energy predict opposite signs of the line tension for the argon-lyophobic solid system.
	}
\end{figure}

\clearpage
\begin{figure}[tb]
	\begin{centering}
		\includegraphics[width=1.0\textwidth]{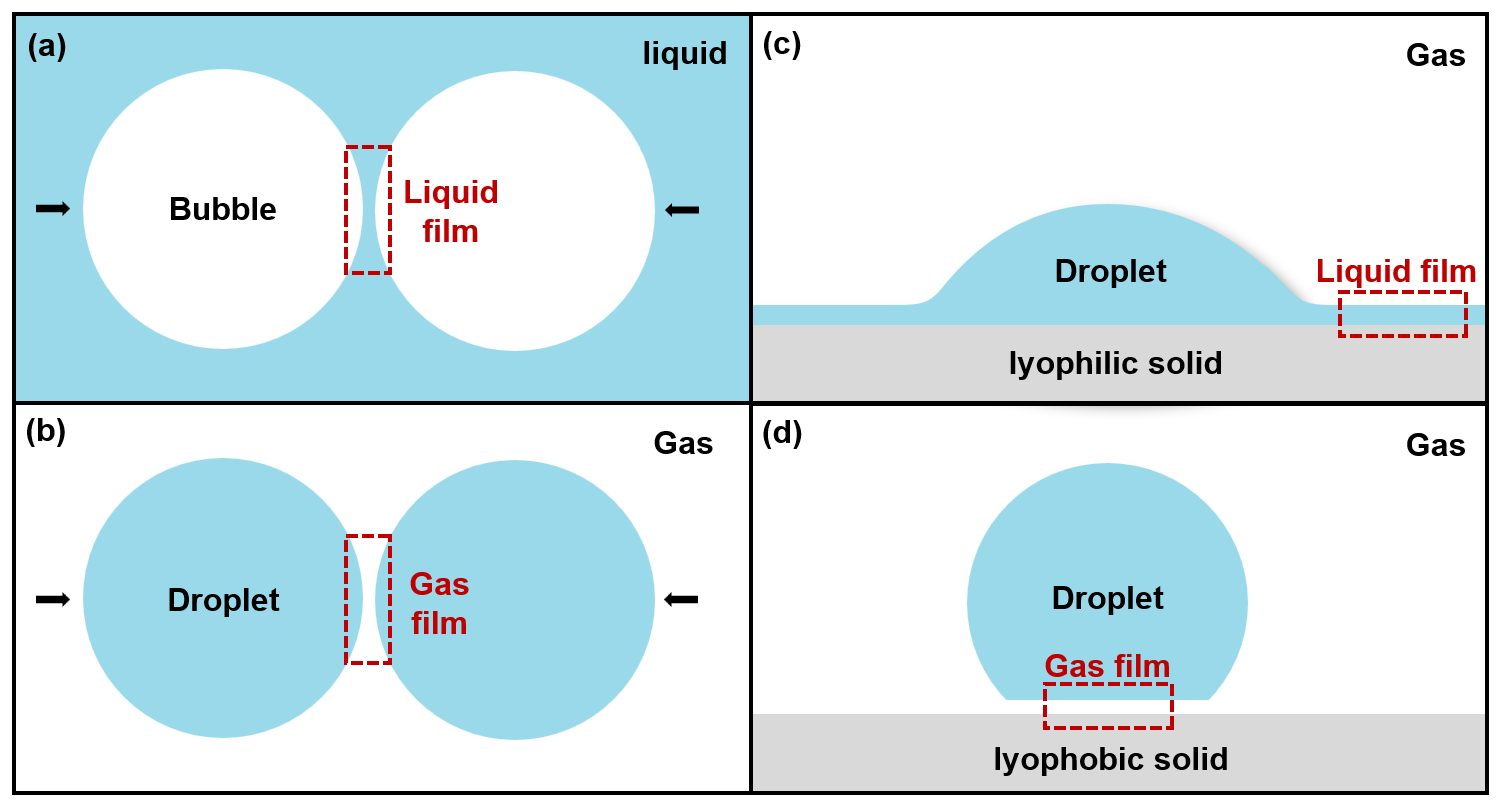}
		\caption{Schematic illustration of fluid nanofilms in different interfacial processes: (a) formation of a liquid-like nanofilm during bubble coalescence; (b) formation of a gas-like nanofilm during droplet coalescence; (c) equilibrium liquid-like nanofilm between the gas phase and a strongly lyophilic surface during droplet wetting; and (d) equilibrium gas-like nanofilm between a droplet and a strongly lyophobic surface.}
		\label{fig:z1}
	\end{centering}
\end{figure}

\clearpage
\begin{figure}[tb]
	\begin{centering}
		\includegraphics[width=0.9\textwidth]{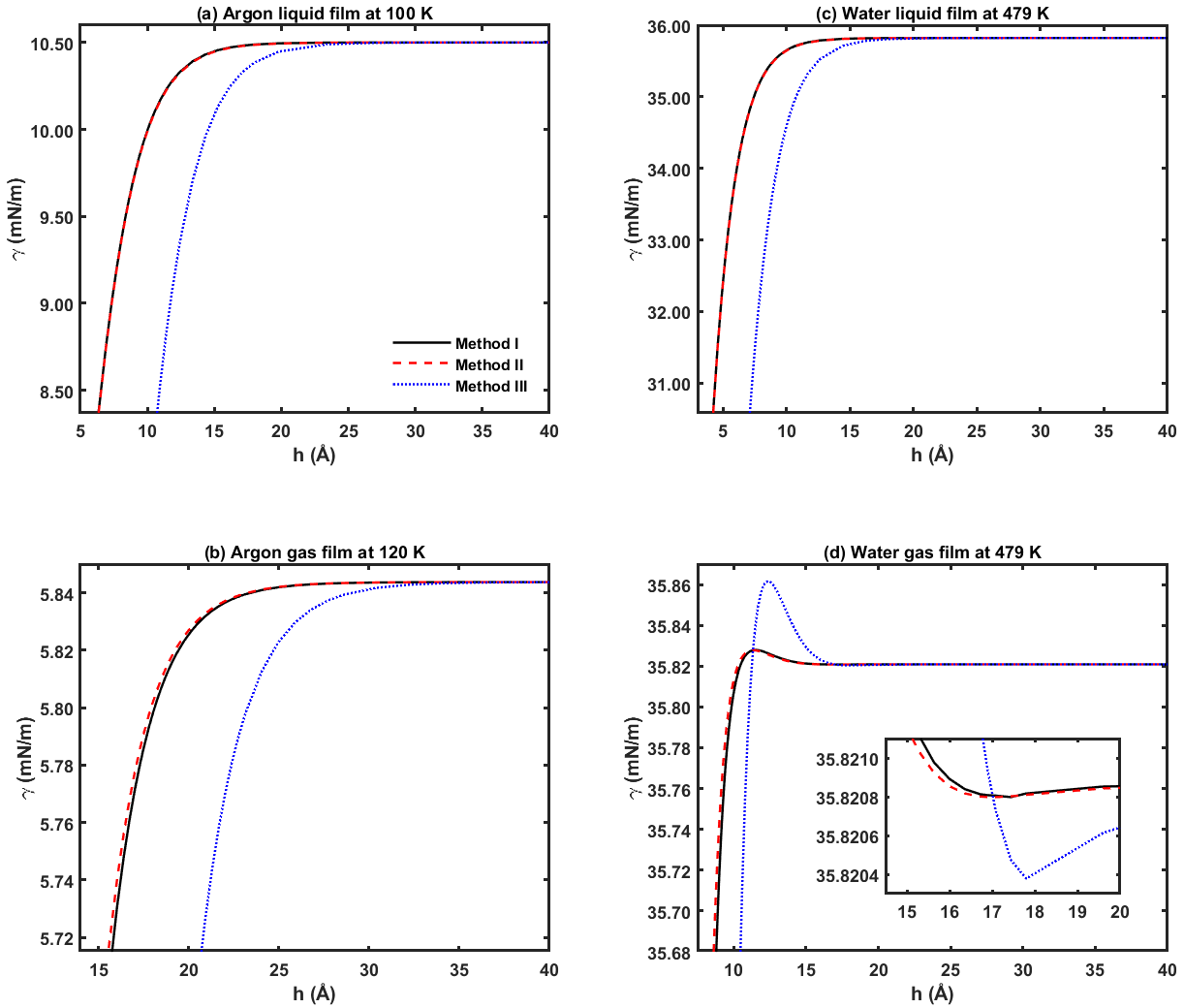}
		\caption{Interfacial free energies (surface tensions) from different methods for liquid-like nanofilms confined by bulk gas phases and gas-like nanofilms confined by bulk liquid phases: (a) argon liquid nanofilm at 100 K; (b) argon gas nanofilm at 120 K; (c) water liquid nanofilm at 479 K; and (d) water gas nanofilm at 479 K. 
		}
		\label{fig:z2}
	\end{centering}
\end{figure}

\clearpage
\begin{figure}[tb]
	\begin{centering}
		\includegraphics[width=0.9\textwidth]{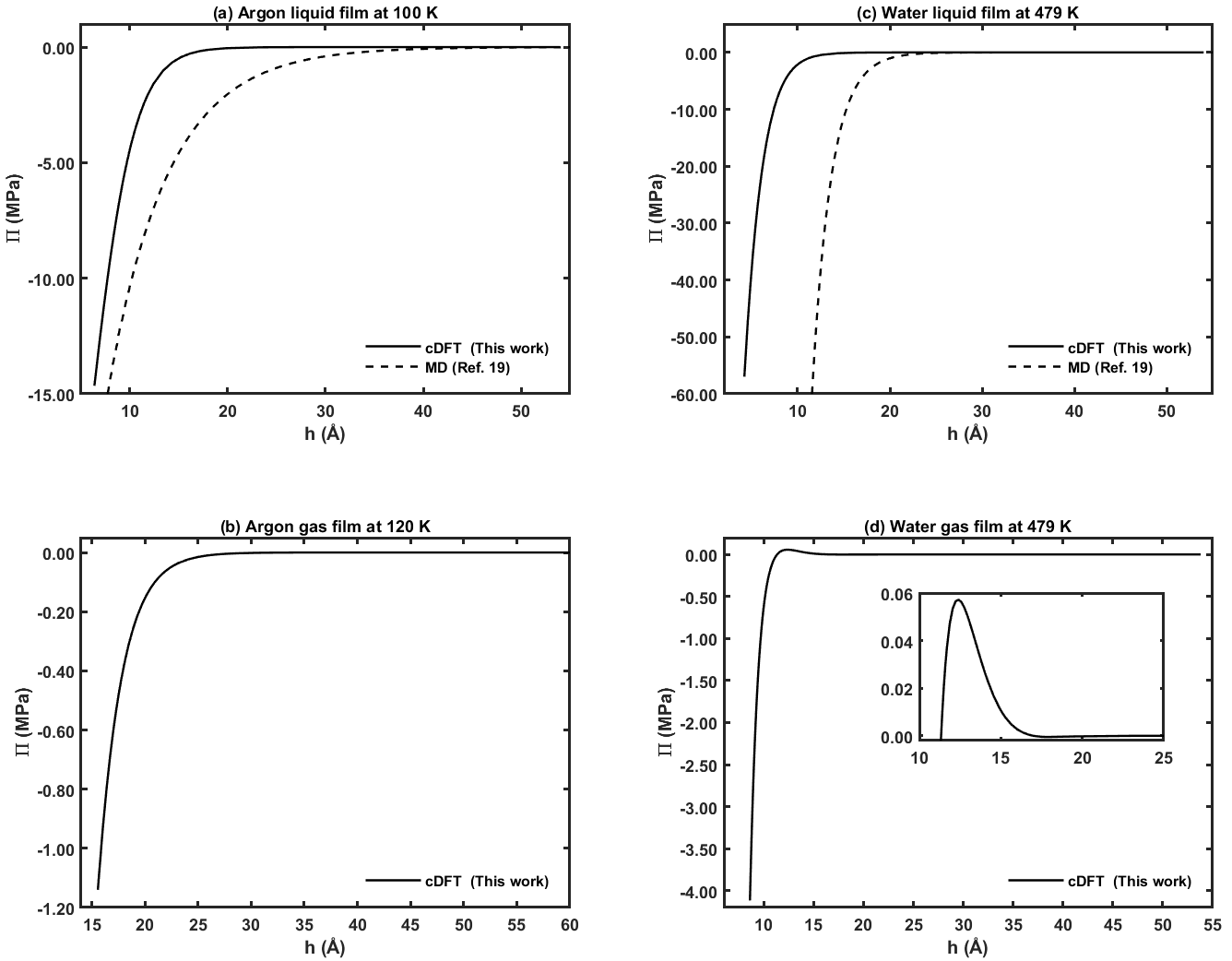}
		\caption{Disjoining pressures of liquid-like nanofilms confined by bulk gas phases and gas-like nanofilms confined by bulk liquid phases: (a) argon liquid nanofilm at 100 K; (b) argon gas nanofilm at 120 K; (c) water liquid nanofilm at 479 K; and (d) water gas nanofilm at 479 K. MD predictions are taken from from Ref. \cite{jianzhou2026estimating}.
		}
		\label{fig:z3}
	\end{centering}
\end{figure}

\clearpage
\begin{figure}[tb]
	\begin{centering}
		\includegraphics[width=0.8\textwidth]{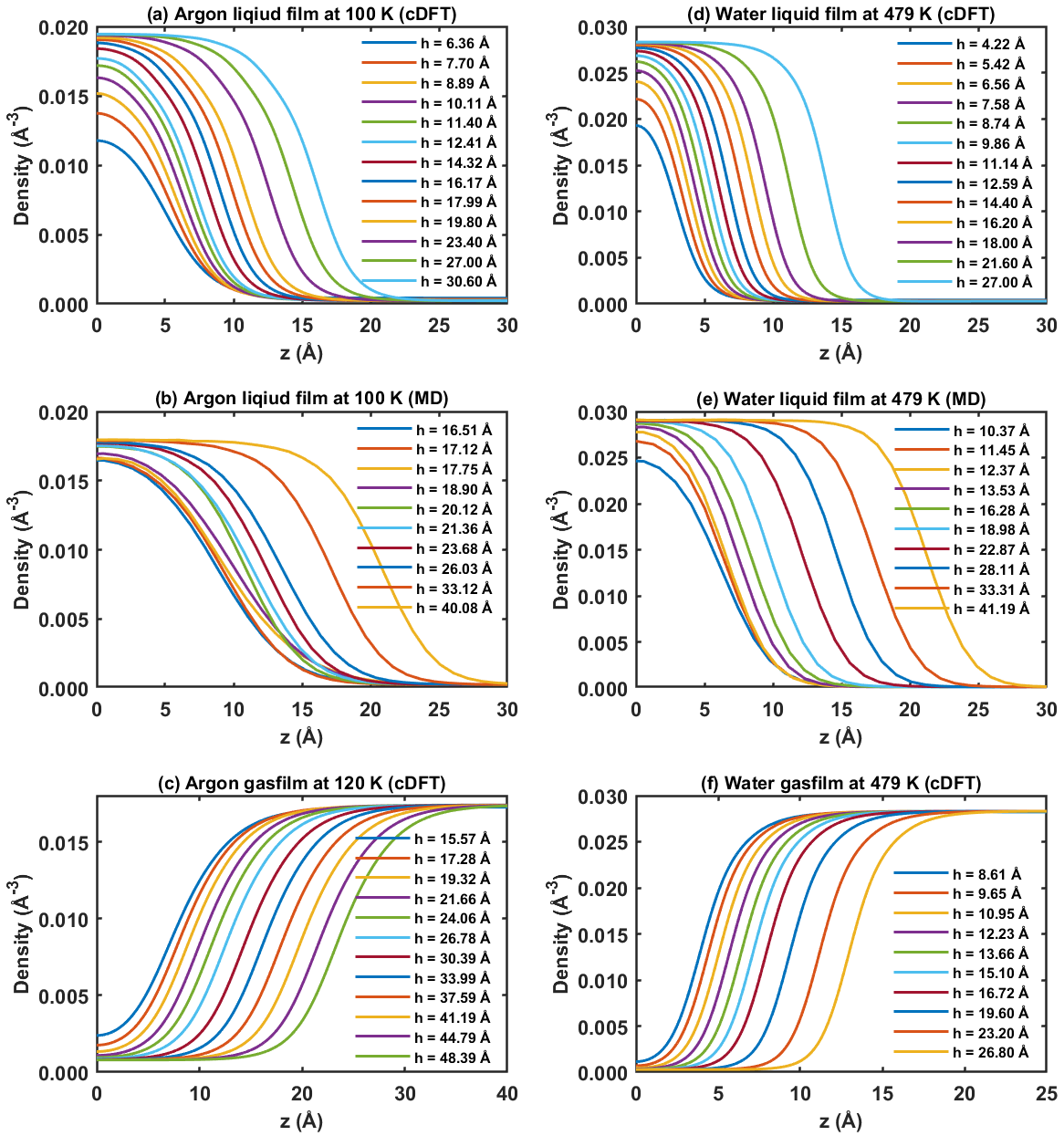}
		\caption{Density profiles of liquid-like and gas-like nanofilms for argon and water at different film thicknesses $h$. Panels (a)-(c) present argon films, including liquid films at 100 K predicted by (a) cDFT and (b) MD, and (c) gas films at 120 K predicted by  cDFT. Panels (d)-(f) present the corresponding results for water, including liquid films at 479 K predicted by (d) cDFT and (e) MD, and (f) gas films at 479 K predicted by cDFT. $z = 0$ corresponds to the midpoint of the film. MD simulation data in (b) and (e) are unpublished data from Ref. \cite{yang2026resolving}. 
		}
		\label{fig:z4}
	\end{centering}
\end{figure}

\clearpage
\begin{figure}[tb]
	\begin{centering}
		\includegraphics[width=0.8\textwidth]{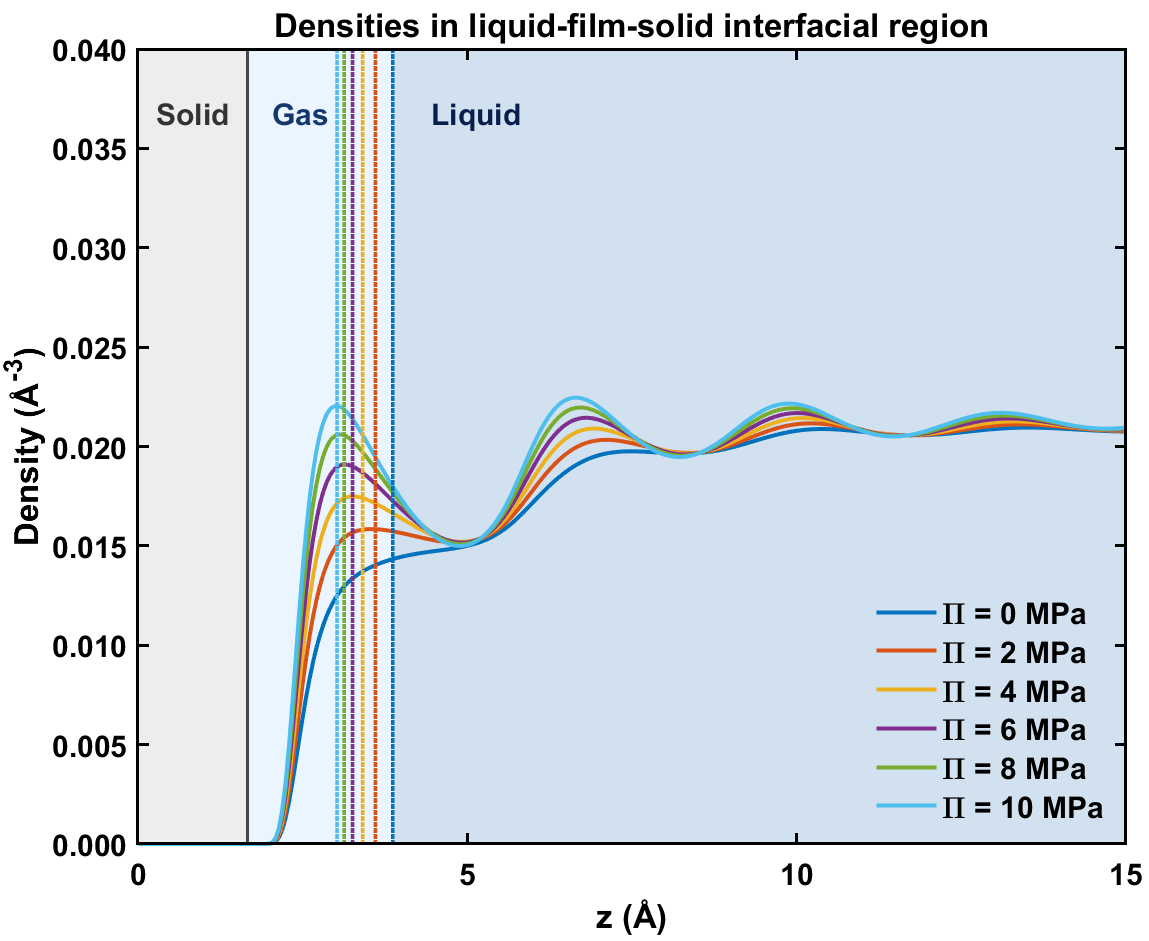}
		\caption{Density profiles of argon in the liquid-film-solid interfacial region at different disjoining pressures. The dotted vertical lines indicate the Gibbs dividing surface for each case. The gray shaded region represents the solid phase. For the $\Pi = 0$ MPa profile, the light blue and dark blue shaded regions indicate the bulk gas and bulk liquid phases, respectively, as defined by the Gibbs dividing surface.
		}
		\label{fig:z5}
	\end{centering}
\end{figure}

\clearpage
\begin{figure}[tb]
	\begin{centering}
		\includegraphics[width=0.8\textwidth]{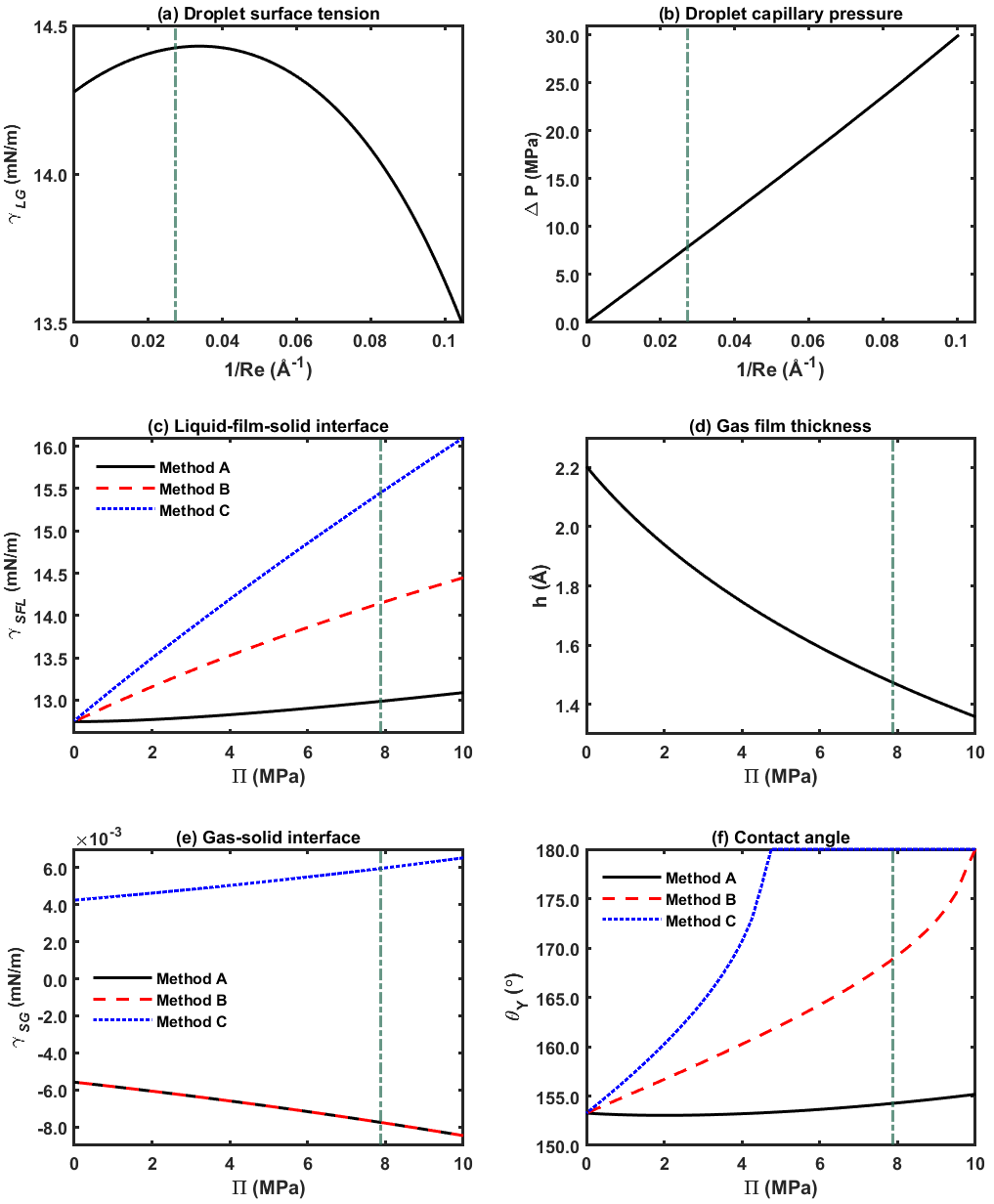}
		\caption{Properties of spherical argon droplets at 83.9 K: (a) surface tension and (b) capillary pressure as functions of the inverse equimolar radius. Properties of hemispherical argon droplets on a lyophobic substrate at 83.9 K as functions of disjoining pressure: (c) liquid-film–solid interfacial free energy calculated using different methods; (d) gas-like film thickness; (e) gas–solid interfacial free energy calculated using different methods; and (f) contact angle $\theta_{\mathrm{Y}}$ predicted by different methods. The green vertical dash-dotted lines indicate the thermodynamic state corresponding to the results shown in Fig. \ref{fig:z7}b.
		}
		\label{fig:z6}
	\end{centering}
\end{figure}

\clearpage
\begin{figure}[tb]
	\begin{centering}
		\includegraphics[width=1.0\textwidth]{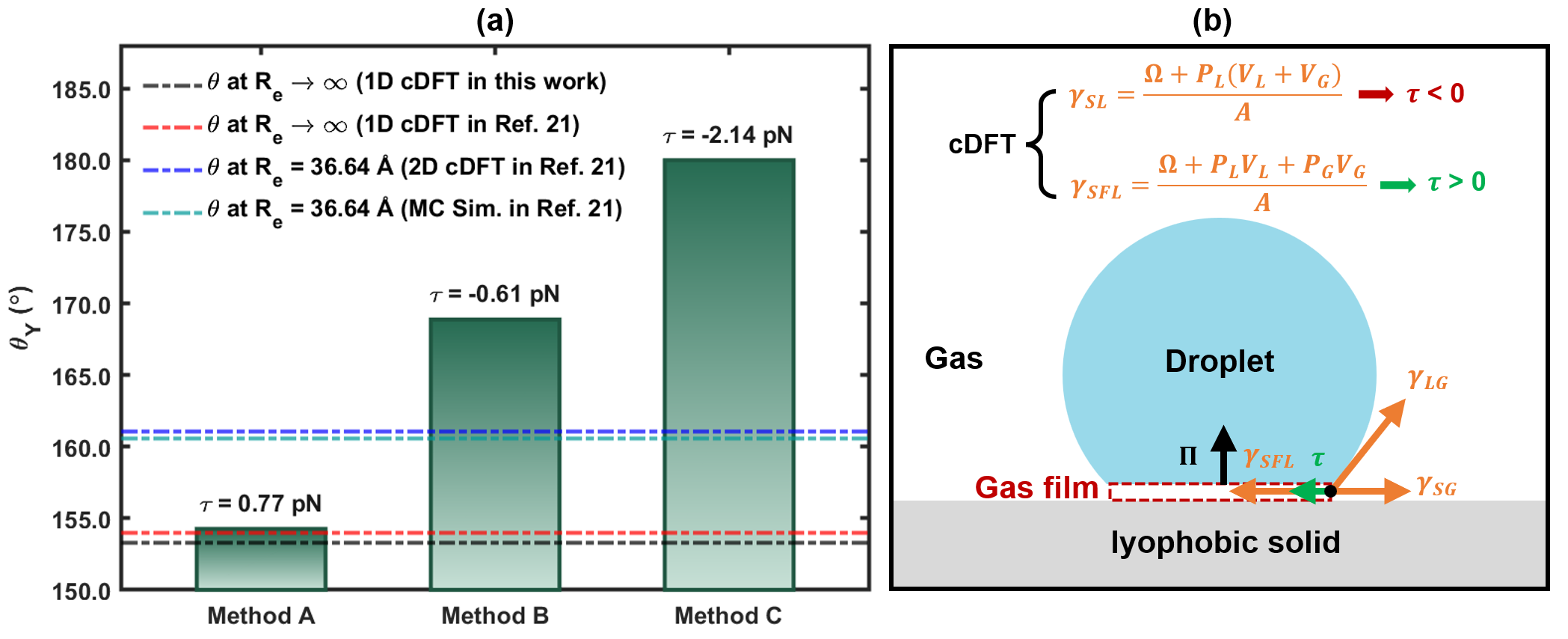}
		\caption{(a) $\theta_{\mathrm{Y}}$ predicted by different methods, compared with $\theta$ from theoretical predictions and simulation data from the literature\cite{sauer2018prediction}, together with the corresponding line tension predicted by Eq. \ref{eq:augYoung2}. The corresponding interfacial properties for this thermodynamic state are marked in Fig. \ref{fig:z6}. (b) Presence of a gas-like nanofilm between a hemispherical droplet and a hydrophobic substrate. Opposite signs of line tension predicted by different methods for droplet-solid IFE in the argon-lyophobic solid system.
		}
		\label{fig:z7}
	\end{centering}
\end{figure}

\end{document}